# Smart Band – An Integrated Device for Emergency Management

A. Jackulin Mahariba, Shivam Patel

*Abstract*: *In the event of a kidnapping or a medical emergency, a person is often incapacitated to be able to call for help. And it's usually too late before the first responders arrive on-scene. Currently, a vast array of 'safety' devices available in the market are often far too rudimentary such as double tapping the power button that isn't very practical, or an app on a smart watch that is a huge investment in the first place. The Smart Band aims to eliminate the need for a physical trigger by the person who finds himself in dangerous situations like kidnapping, front-facing some wild animal, heart attack, etc., and rather senses the heartbeat. The Smart Band is designed as a personalized wearable device, wherein the user heart beat rate is collected and trained using machine learning algorithm, which triggers the alert system automatically when the event is identified. Hence the accuracy of assessing emergency situation will be high and false rate will be reduced. As soon as the event is detected, the band relays GPS coordinates to first responders and emergency contacts, which will be sent via the Network Carrier (SIM card module) directly from the band, not relying on a mobile phone, which is usually out of reach during such emergency situations. In essence, the Smart Band consists of a GPS tracker, a heartbeat sensor, a Network module, and a Bluetooth module, all existing technologies which have been mass produced to an extent that the end product can be made affordable, and in huge quantities as well. On further development the smart band can be customized to a finely wearable jewel which can serve the purpose autonomously.*

## I. INTRODUCTION

According to the comprehensive legislation of India, the offenses for kidnapping are imposed based on the reason or motive behind the kidnapping. The purposes of kidnapping as per IPC are for begging, for ransom, in order to murder, intent to confine a person, importation of girl child from other country, child slavery, women to compel for a marriage and so on. Every purpose or intention behind kidnapping have different section of penal code with its own penalty and judgement. Though we have strict rules and regulations against such sort of activities, the crime records are not reducing. The National Crime Records Bureau (NCRB) shows that their share in total crimes against women nearly doubled from 10% in 2001 to 19% in 2016. According to the 2017-18 report of the MHA, 54,723 children were kidnapped in 2016 but charge sheets were filed in only 40.4% of the cases [5]. The MHA report also revealed that 8,132 cases of human trafficking were registered in the country in 2016.Whenever such incident happen to any human being irrespective of gender and age, it is equally important to safeguard them against the kidnappers. Though there are different purpose lying behind any such incident, the final condition of the victim on majority of the cases results in loss of life. There are many recent advancements in technologies helps us in protecting wealth and treasures against theft and disaster. But for cause of saving a life is not cared much throughout the world. It motivates us in proposing a system which can automatically initiate the rescue operation of the victim from any uncomfortable situation caused by the kidnappers to the hostages. The apprehensive behavior of any human being will cause stress in them. The stress will induce the irregular heartbeat. Such irregularity is called as arterial fibrillation. It is defined as the state of having rapid, skipped and fluttering heartbeat, hence the pulse rate will be faster than normal.

Carrying on restlessly actuates the pressure reaction. The pressure reaction promptly causes explicit physiological, mental, and enthusiastic changes in the body that upgrade the body's capacity to manage a danger—to either battle with or escape from it—which is the reason the pressure reaction is regularly alluded to as the battle or flight reaction. Some portion of the pressure reaction changes incorporate animating the pulse so as to roundabout blood all through the body so it is better outfitted to manage a danger.

At the point when stress reactions happen rarely, the body can regain generally rapidly from the physiological, mental, and passionate changes the pressure reaction realizes. At the point when stress reactions happen too regularly or potentially drastically, be that as it may, the body has an increasingly troublesome time recouping, which can result in the body staying in a semi hyper stimulated state, since stress hormones are stimulants. A body that progresses toward becoming pressure reaction hyper stimulated can display comparative sensations and manifestations to that of a functioning pressure reaction.

## II. RELATED WORK

There is proof to recommend that how best to help the individuals who have been abducted is sensitive furthermore, complex issue, and the individuals who manage such people ought to be also educated as would be prudent since such occasions can have long term adverse results, especially on kids [1]. The post effects on such cases will cause cognitive, emotional and social imbalance in the victims.

The ECG recordings are used identify arrhythmias efficiently, which serves as a tool for cardiologists. A multinomial classification for different types of heart abnormalities is implemented using decision tree algorithm

**A. Jackulin Mahariba**, Department of Computer Science and Engineering, SRM Institute of Science and Technology, Kattankulathur, Chennai, Tamil Nadu, India. E-mail: jackulin.a@ktr.srmuniv.ac.in

**Shivam Patel**, Department of Computer Science and Engineering, SRM Institute of Science and Technology, Kattankulathur, Chennai, Tamil Nadu, India. E-mail: shivam.patel1606@gmail.com

on the UCI machine learning repository containing an arrhythmia dataset [2]. The inflection point is detected in ECG signals and the heart rate is calculated. Then SVM classifier is applied on the heart rate to classify the results into two classes as 0 and 1. The classification results in normal or abnormal in the heart rate [3].

Popular data mining algorithms such as Naïve Bayes, Artificial Neural Networks, K-Means, and Decision Tree are implemented on a dataset to create models to infer and draw conclusions based on the results of such algorithms. [4] shows that Naïve Bayes performs better than the other algorithm.

## III. PROPOSED METHODOLOGY

From the studies it is strongly evident that the stress caused to the human being will cause the sudden change in the pulse rate in the form of arterial fibrillation. The victim will be in a state of isolation and unable to communicate to the world. The proposed system will automatically check for the stress induced arterial fibrillation and trigger the event of conveying the dangerous situation of the person to the emergency contact numbers and to the help line number along with the GPS location of the victim. The system consists of four major components with a processor. They are pulse rate sensor, detection unit, GPS transceiver and GSM module.

The architecture diagram of the proposed system is given in the Fig. 1. The pulse rate sensor is used to continuously monitor the pulse rate to check for the abnormal rise or low in the sensor output. Optical pulse sensors utilize a procedure called photoplethysmography (PPG) to quantify pulse [6]. PPG is a specialized term for sparkling light into the skin and estimating the measure of light that is dissipated by blood stream. That is a touch of a distortion, yet PPG depends on the way that light entering the body will dissipate in an anticipated way as the blood stream elements change, for example, with changes in blood beat rates (pulse) or with changes in blood volume (cardiovascular yield).

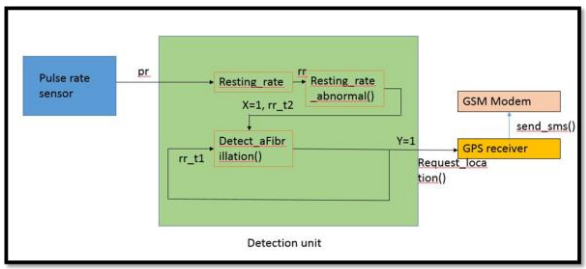

**Fig. 1: Architecture Diagram for the Design of Smart Band**

Many smart phone companies have launched smart devices like smart watches, wrist bands and fitness bands which helps to monitor the pulse rate abnormality of the patients and common people. It detects the abnormality and sends notification to their mobile phone in the form of alert message. The major drawback of such devices are such health monitoring devices are designed to react in general with reference to the standard pulse rate measures. But the resting rate of any person need not be unique with reference to standard rates. Hence in our system we plan to utilize the machine learning algorithms to customize the function of health monitoring since it varies from individual to individual. The pulse rate is collected initially all form of activities in daily life (ADL) such as walking, jumping, hopping, swimming, during exercise, sleeping and so on. The variation of pulse rate at ADL are shown in the Fig. 2.

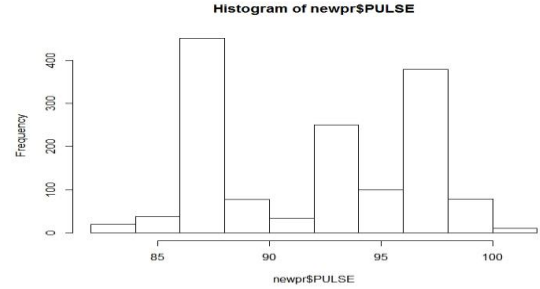

**Fig. 2: Histogram shows the peaks in pulse during ADL**

Then the regression model is developed to find the resting rate (rr) of the individual using the pulse rate. Through which the threshold (τr) is determined by calculating the mean of the resting rate.

The fluttering in the resting rate causes the Arterial fibrillation can also be called as panic attack. There are few differences between the panic attack and heart attack. In the panic attack the rapid rise or slowdown of pulse rate occur at a time, then the heart pumping will be regular once the victim is emotionally stable. But in case of heart attack the stabilization will not occur till the doctor treats the patient. Whatever might be the reason for palpitation, both are dangerous situation to handle by any human by themselves. Hence the detection of the rapid change in the pulse rate is monitored to verify the livelihood condition of the victim. Once the abnormality is encountered then the system will automatically send a notification message along with the location detail in the form of a map.

## IV. IMPLEMENTATION

The pulse rate sensed by the sensor is fed as the input to the detection unit. There are two primary conditions to be checked to detect the dangerous condition of the victim. First condition is to detect abnormality in the heart rate. The detection unit takes the pulse rate sensor output and calculate the resting rate of the person through the model developed using linear regression. The publicly available dataset in physio net [7] is used for the results presented in this paper. The data set contains 4 features they are, heart rate, pulse rate, resting rate and SPO2. The linear regression model is developed on the data set to predict the resting rate of the person using the pulse rate obtained from the pulse rate sensor.

Coefficients:   Intercept   newpr$PULSE
                41.1532     -0.2886

$$rr = 41.1532 - 0.2886 * pr$$

The resting rate (rr) is compared with the threshold of the resting rate (τr) of the victim. The rise or fall in the resting rate is considered as abnormality. The second condition is checked to confirm the occurrence of arterial fibrillation. The detection unit consists of a feedback loop where the pulse rate of rr_t1 and rr_t2 are compared to verify the fluttering in the heart rate. This feedback loop can be simple flip-flop to check the variation in the rr.

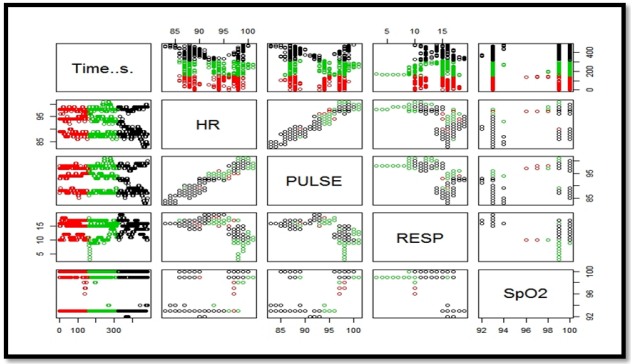

**Fig. 3: Features in the Representation Dataset plotted pairwise from Physionet**

The pair plot (Fig. 3) with respect to all the parameters in the dataset the variation in the pulse rate is grouped in two three clusters with high variations. The data distribution of the four features is observed, in which the resting rate has the outlier refer Fig. 4. The k means clustering algorithm is applied on the dataset with center 3 results in the variability of 79.31% while with the center 5 results in 75.16%, and the same is plotted in the Fig. 5 and 6.

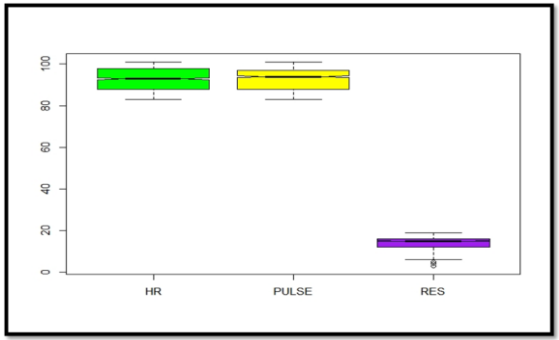

**Fig. 4: Distribution of data on features – HR, PULSE, and RES**

The device is now trained with all the variations (maximum and minimum range) in the pulse rate of the user with all the ADL. And the device is now customized for the user's activity. If the deviation from the already observed pattern occurs then the device will be activated for the emergency management.

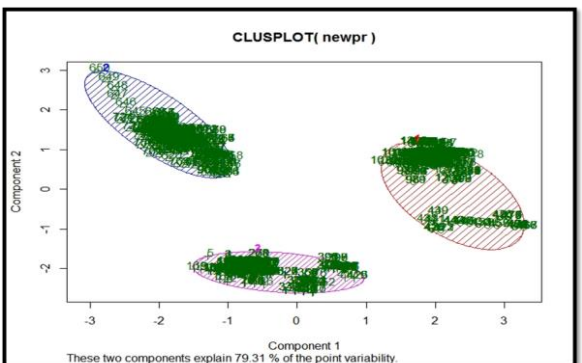

**Fig. 5: Clusters formed on applying clustering algorithm with center 3**

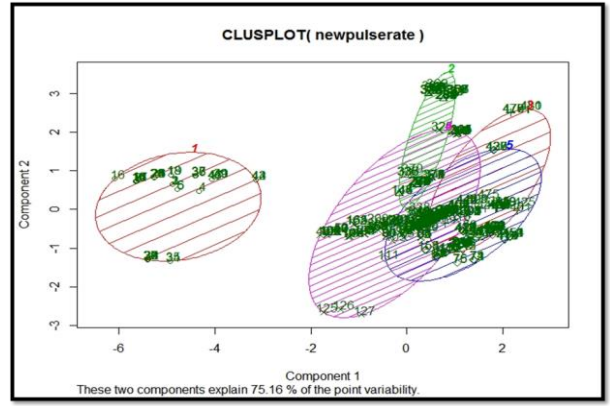

**Fig. 6: Clusters formed on applying clustering algorithm with center 5**

## V. CONCLUSION AND FUTURE ENHAMCEMENT

As the number of centers are increased the variability decreases, which will degrade the performance of fixing the resting rate. Hence the accuracy of assessing emergency situation will be high and false rate will be reduced. As soon as the event is detected, the band relays GPS coordinates to first responders and emergency contacts, which will be sent via the Network Carrier (SIM card module) directly from the band, not relying on a mobile phone, which is usually out of reach during such emergency situations. On further development the smart band can be customized to a finely wearable jewel which can serve the purpose autonomously.